# Generate your neural signals from mine: individual-to-individual EEG converters

**Zitong Lu (lu.2637@osu.edu)**
Department of Psychology, The Ohio State University, 1827 Neil Ave
Columbus, OH 43212 USA

**Julie D. Golomb (golomb.9@osu.edu)**
Department of Psychology, The Ohio State University, 1827 Neil Ave
Columbus, OH 43212 USA

**Abstract**

Most models in cognitive and computational neuroscience trained on one subject don't generalize to other subjects due to individual differences. An ideal individual-to-individual neural converter is expected to generate real neural signals of one subject from those of another one, which can overcome the problem of individual differences for cognitive and computational models. In this study, we propose a novel individual-to-individual EEG converter, called EEG2EEG, inspired by generative models in computer vision. We applied THINGS EEG2 dataset to train and test 72 independent EEG2EEG models corresponding to 72 pairs across 9 subjects. Our results demonstrate that EEG2EEG is able to effectively learn the mapping of neural representations in EEG signals from one subject to another and achieve high conversion performance. Additionally, the generated EEG signals contain clearer representations of visual information than that can be obtained from real data. This method establishes a novel and state-of-the-art framework for neural conversion of EEG signals, which can realize a flexible and high-performance mapping from individual to individual and provide insight for both neural engineering and cognitive neuroscience.

*Code:* https://github.com/ZitongLu1996/EEG2EEG

## Introduction

Individual differences play a significant role in cognitive and computational neuroscience, as neural activity varies greatly from person to person. This variation can make it difficult to generalize models trained on one subject to other subjects. However, if a model can predict the neural signals of one individual from those of others, it can not only help reduce the cost of experiments but also provide an opportunity to better understand the mechanism of individual differences. Therefore, it is important to consider the development of a neural signal generator that can overcome these individual differences across individuals.

Functional MRI (fMRI) researchers often use normalization techniques, such as aligning different individuals' brains to a common template (Evans et al., 1992, 1994; Hasson et al., 2004), when tring to generalize across individuals. However, anatomical alignment cannot align the functional topography across individuals well (Watson et al., 1993; Yamada et al., 2015). Additionally, this approach is not feasible for low spatial resolution recordings, such as electroencephalography (EEG), which contains much more temporal information. This highlights the need of alternative methods to overcome the problem of individual differences in neural activity, particularly for EEG.

Recently, an anatomy-free method called functional alignment has been used to directly learn the functional differences across different subjects' brain activity (Bilenko & Gallant, 2016; Chen et al., 2015; Gu et al., 2022; Guntupalli et al., 2016; Haxby et al., 2011). Most works of functional alignment constructed a common space of neural representations by inputting fMRI activity across individuals to achieve brain activity conversions. However, this approach has limitations such as the need for a large number of subjects to obtain an accurate common space, and the spatial locations of EEG channels across sessions and subjects not being as stable as MRI voxels. Also, we expect to denoise the data which allows for more clear representational patterns. Is it possible to propose a more flexible, individual-to-individual, EEG converter that addresses these limitations?

Several studies have attempted to use a linear model as a neural code converter (Ho et al., 2023; Yamada et al., 2015) to learn the mapping of fMRI signals from a source subject to a target subject. However, there is a lack of studies in EEG conversion. The EEG converter based on a linear model, as a baseline model, may not be appropriate because EEG has a lower signal-to-noise ratio than fMRI and contains complex timing information. But in the field of computer vision, a lot of generative models with high performance at different tasks have emerged in recent years, which can provide insights of building novel models for pairwise EEG conversion.

In this study, we propose a novel individual-to-individual EEG converter called EEG2EEG. Our EEG2EEG converter, inspired by U-Net and its applications (Çiçek et al., 2016; Isola et al., 2017; Kandel et al., 2020; Ronneberger et al., 2015; Yao et al., 2018), is a nonlinear model with full 1-D convolutions. It can learn the mapping of EEG signals from a source subject to a target subject and generate realistic EEG signals from unseen images. We not only evaluated our model from the perspective of data similarity between real and generated signals, but also verified that the generated EEG signals based on EEG2EEG contained sufficient visual information of low-level to high-level visual features from the perspective of information representation. This pairwise EEG conversion framework provides great feasibility for the generation of individualized neural signals in the future and also offers a new perspective for investigating the mechanism of individual differences.

## Methods

**EEG dataset**
The EEG recordings used in this study were obtained from an EEG open dataset, THINGS EEG2 (Gifford et al., 2022). The dataset consists of EEG responses from ten participants while they were performing a rapid serial visual presentation (RSVP) experiment. Each participants completed a total of 82,160 trials spanning 16,740 image conditions. This dataset was divided into a training dataset (16,540 training images with 4 repeated trials) and a test dataset (200 test images with 80 repeated trials). Additional information about the image stimuli and EEG recordings can be found in the reference provided (Gifford et al., 2022).

**Data preprocessing**
We epoched the continuous EEG data into trials ranging from stimulus onset to 200ms after stimulus onset based on nine subjects' raw data with a sample frequency of 1000Hz. The EEG data was baseline corrected by subtracting the mean voltage from 50ms before stimulus onset to stimulus onset. One subject's data (Sub 09) was excluded due to the lack of one session's data. We selected 17 channels from the occipital and parietal areas for further analysis (O1, Oz, O2, PO7, PO3, POz, PO4, PO8, P7, P5, P3, P1, Pz, P2, P4, P6, P8). For training dataset, 4 trials corresponding to the same image were averaged to obtain cleaner EEG signals. The matrix of training data for each subject was 16540 images × 17 channels × 200 time-points. For test dataset, we used two options. For the ERP level test set, 80 trials corresponding to the same image were averaged, resulting in a test data matrix for each subject of 200 images × 17 channels × 200 time-points. For the single-trial level test set, trials were not averaged, and the matrix of the test data for each subject was 16000 images × 17 channels × 200 time-points.

**EEG2EEG Model**
The goal of EEG2EEG converter $C$ is to learn a mapping from a source subject's EEG signal $s$ to a target subject's EEG signal $t$, $C: s \rightarrow t$. EEG2EEG is trained to produce generated EEG signals that are as similar as possible to the real target subject's EEG signals.

**Objective** Given training pairs $\{(s, t)\}$, the objective of a EEG2EEG can be expressed as
$$\mathcal{L}^C = \mathcal{L}_t(\hat{t}, t)$$

$\mathcal{L}^C$ consists of losses on EEG amplitudes measuring the mean squared error (squared L2 norm), $\mathcal{L}_{MSE}(\hat{t}, t)$, and EEG patterns measuring the cosine distance, $\mathcal{L}_{cos}(\hat{t}, t)$. The loss for a generated EEG signal $\hat{t}$ reads:
$$\mathcal{L}_t(\hat{t}, t) = \mathcal{L}_{MSE}(\hat{t}, t) + \mathcal{L}_{cos}(\hat{t}, t)$$
$$\mathcal{L}_{MSE}(\hat{t}, t) = \frac{1}{N_{ch}N_{tp}} \sum_i^{N_{ch}} \sum_j^{N_{tp}} (\hat{t}_{i,j} - t_{i,j})^2$$
$$\mathcal{L}_{cos}(\hat{t}, t) = 1 - \cos(\hat{t}, t)$$
where $N_{ch}$ and $N_{tp}$ are the numbers of channels and time-points. $\cos(\hat{t}, t)$ denotes the cosine similarity between generated and real signals.

**Network architecture** The network architecture is illustrated in Figure 1A, which is modified based on the structure of the original U-Net (Çiçek et al., 2016). The network is composed of two main parts, an encoder (contracting path, left side) and a decoder (expansive path, right side). The encoder consists of the repeated application of two 1D convolutions with kernel size of 3, each followed by a rectified linear unit (ReLU) and a max pooling operation with kernel size of 2 and stride of 2 for downsampling. The decoder consists of an upsampling of the feature map, followed by a 1D up-convolution with kernel size of 3, a concatenation with the correspondingly feature map from the encoder layer, two 1D up-convolutions with kernel size of 3, and a ReLU. At the final layer, a 1D convolution with kernel size of 1 is applied to match the output to the shape of the EEG matrix. In total, EEG2EEG model has 8 layers. The skip connections between each layer $i$ and layer 8-$i$ help the decoder to generate more typical EEG patterns. All the 1D convolutions in EEG2EEG are applied independently for each channel to account for the unknown relationship between different channels in EEG data. We initialized the weights using normal initializer.

**Model training and test** We aimed to train 72 neural converters, each corresponding to a pair of one source subject and one target subject. For each pair, we first normalized the amplitudes of the source subject's EEG data, and then applied that normalization to the target subject's EEG data before training. During the training process, we applied batch normalization and trained each EEG2EEG using Adam optimizer for 100 epochs with a learning rate of 0.002 and batch size of 32. For ERP level test, we input the data after averaging repeated trials. For single-trial level test, we directly input the test data from the source subject and then average every 80 trials.

**Performance evaluation** To evaluate how generated data based on a EEG2EEG converter was similar to real target data, we first calculated the overall generated accuracy by calculating the Spearman correlation coefficient between the generated data array and real target data array (e.g., ERP-level: 200 images × 17 channels × 200 timepoints). Secondly, we calculated the pattern-generated accuracy by calculating the correlation between the generated and real data for each image and then averaging the correlations across all images. Thirdly, we calculated the profile correlation by calculating the correlation for each time-point and then averaging the correlations across all time-points. We evaluated the generated accuracy at both ERP and single-trial levels.

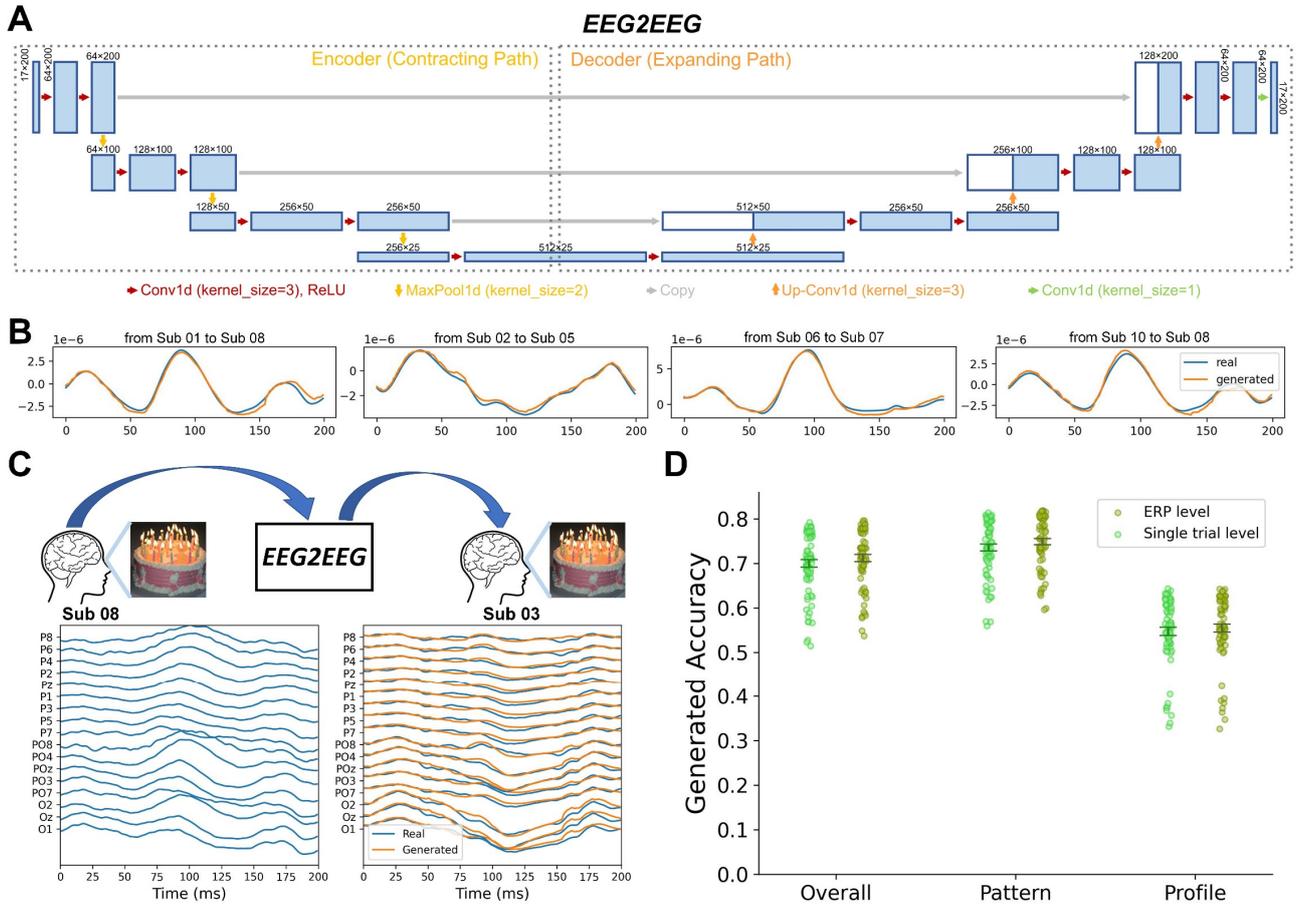

Figure 1: (A) Structure of EEG2EEG. (B) Four examples of pairwise conversion results that we averaged all channels and all image conditions. (C) An example of EEG conversion from Sub 08 to Sub 03 corresponding to a single image condition. (D) Overall, pattern, and profile generated accuracies at ERP and single-trial levels. Error bars reflect ± SEM.

**Correlation analyses** To explore the relationships between EEG2EEG generated accuracy and other factors, we calculated the Pearson correlation coefficients between the generated accuracy from subject A to subject B and the accuracy from subject B to subject A; between the generated accuracy and EEG similarity between two subjects; and between the generated accuracy and EEG variability of both target and source subjects respectively.

**Multivariate pattern analysis (MVPA)**

In addition to evaluating the similarity between generated data and real target data, we also aimed to assess whether generated data based on EEG2EEG contains the same quality of visual information as real data. To do this, we applied two MVPA methods, classification-based decoding (Grootswagers et al., 2017; Schaefer et al., 2011) and representational similarity analysis (Kriegeskorte et al., 2008), to analyze the neural representations in both generated and real data. This allowed us to evaluate the information content of the two types of data.

**Classification-based decoding** We applied linear support vector machine (SVM) to classify the neural patterns of each pair of images on single-trial level test data for each 10ms time-window with a 10ms time-step. The classifier was trained and tested independently for each time-window. To improve the signal-to-noise ratio, we averaged every 5 trials from same image condition. We evaluated each classifier using a 5 × 4-fold cross-validation framework. The decoding accuracies were averaged across all image pairs and compared to a chance level of 50%. A higher decoding accuracy suggested a greater difference between neural patterns. For real data, the decoding analyses were performed separately for every subject. For generated data, the decoding analyses were performed separately for every target subject from all 72 EEG2EEG converters.

**Representational similarity analysis (RSA)** Previous studies have suggested that layers in convolutional neural networks (DCNNs) capture neural representations of visual information from low-level to high-level (Cichy et al., 2016; Güçlü & van Gerven, 2015; Kietzmann et al., 2019; Yamins et al., 2014). We compared the representations of 200 test images between human brains and a pretrained DCNN by calculating representational dissimilarity matrices (RDMs) of EEG and DCNN and conducting the comparisons. The shape of each RDM was 200 × 200. For EEG RDMs, we applied decoding accuracy between two image conditions as the dissimilarity index to construct EEG RDM for each time-window and each subject. We obtained 9 sets of 20 RDMs based on real data and 72 sets of 20 RDMs based on generated data (every 10ms as a time-window to calculate the RDM). For DCNN RDMs, we input all 200 images into an AlexNet pretrained on ImageNet and extracted features from each layer. Then we calculated one minus the Pearson correlation coefficient between the activation vectors corresponding to any two images as the dissimilarity index in the RDM for each layer. In addition, we also calculated a low-level RDMs directly based on original images. Thus, we obtained 8 RDMs corresponding to 8 layers in AlexNet and 1 low-level RDM. To compare the representations, we calculated the Spearman correlation coefficient as the similarity between the EEG RDMs and DCNN RDMs. Both decoding and RSA analyses were implemented based on NeuroRA toolbox (Lu & Ku, 2020).

## Results

**EEG2EEG model performance**
Using EEG2EEG model as the neural converter, we could overcome the individual different to generate very realistic EEG signals of a target subject from a source subject. Figure 1B shows some examples of pairwise EEG2EEG conversion results averaging across all trials and all channels. The close correspondence between the real EEG data from a given target subject (blue lines) and the generated EEG signal of that target subject from a different source subject (orange lines) suggests that generated EEG signals were very similar to the real EEG signals.

Additionally, Figure 1C presents an example of EEG signals corresponding to a single image condition from Sub 08 as the source subject to Sub 03 as the target subject, when they observed the same image of a birthday cake. Despite the original signals (blue lines) from Sub 08 and Sub 03 were very different due to individual differences, the EEG2EEG converter was able to learn the mapping from Sub 08 to Sub 03, and generate EEG signals (orange lines) that were quite similar to Sub 03's real EEG signals.

To quantify the model performance, Figure 1D shows generated accuracies at both ERP and single-trial level. Overall accuracies were 71.2% ± 6.7% at ERP level and 70.0% ± 7.1% at single-trial level, and the pattern accuracies were 74.9% ± 6.0% at ERP level and 73.6% ± 6.4% at single-trial level. Although profile accuracies were not as high as pattern accuracies, they were still 55.5% ± 7.6% at ERP level and 54.7% ± 7.7% at single-trial level. These results suggest that EEG2EEG models can achieve very high conversion performance of EEG signals from one subject to another.

**Correlations between generated accuracy and other factors**

To investigate which factors could influence generated accuracy of EEG2EEG, we conducted several correlation analyses between generated accuracy and other factors. Firstly, there was no significant correlation between the generated accuracy from subject A to subject B and the accuracy from subject B to subject A (Figure 2A). Secondly, there was no significant correlation found between the generated accuracy and EEG similarity between two subjects (Figure 2B). Thirdly, there was no significant correlation between the generated accuracy and EEG variability of the *source* subject, but there was a significant correlation between the generated accuracy and EEG variability of the *target* subject (Figure 2C). Here, we calculated EEG variability using the inter-trial cosine distance corresponding to the same image condition based on the training data. These results suggest that EEG2EEG conversion performance is primarily influenced by the EEG variability of the target subject.

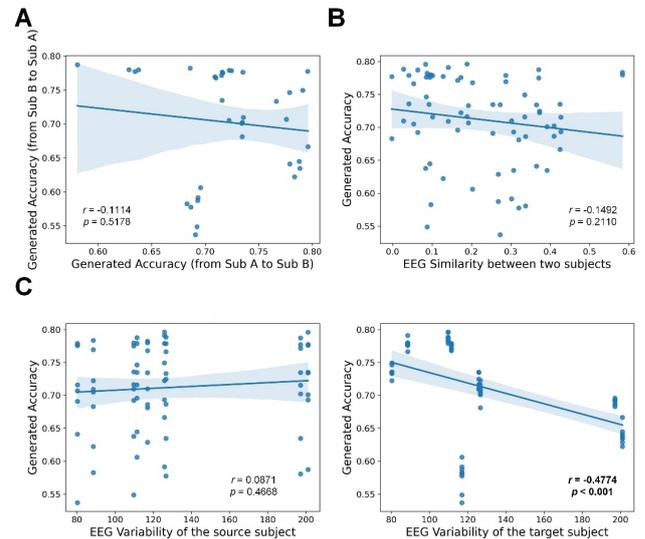

Figure 2: Correlation analyses. (A) The correlation between the generated accuracy from subject A to subject B and the accuracy from subject B to subject A. (B) The correlation between the generated accuracy and EEG similarity between two subjects. (C) The correlation between the generated accuracy and EEG variability of the target subject and the source subject. Ribbons represent the 95% CI. Smaller dots represent individual data points of 72 pairs.

### Comparing EEG2EEG model to others

We conducted ablation experiments to evaluate the effectiveness of key components in our EEG2EEG model. We trained two additional models, one that had no skip connection (Non-Connection) and one that had no cosine loss function (Non-CosineLoss) in original EEG2EEG architecture. In addition, we trained a full-connected linear model that directly connected input and output EEG data from source and target subjects as a baseline. Table 1 shows the mean and standard deviation of generated accuracies of 72 pairs. The results show that all three nonlinear models had higher generated accuracies than the linear model, with the full EEG2EEG model achieving the highest performance. These results indicate that both key factors, skip connections and cosine loss function, can significantly improve the performance of the neural converter.

Table 1: Generated accuracy comparison in ablation study.

| Method | ERP level | single-trial level |
| --- | --- | --- |
| Linear model | 0.640 ± 0.095 | 0.640 ± 0.095 |
| Non-Connection | 0.676 ± 0.158 | 0.665 ± 0.157 |
| Non-CosineLoss | 0.709 ± 0.069 | 0.697 ± 0.072 |
| EEG2EEG | **0.712 ± 0.067** | **0.700 ± 0.071** |

### Information representation in real and generated EEG signals

Figure 3 shows decoding accuracy and representational similarity of real EEG data and generated EEG data. In Figure 3A-B, the mean decoding accuracy of image identity over time is presented for both real data from 9 subjects and generated data based on EEG2EEG from 72 pairs. It was found that the decoding accuracy for real data was significantly above chance level starting from 50ms after image onset, while generated data was significant from 20ms. Additionally, decoding accuracies based on generated data were higher than those from real data.

Figure 3C-D show representational similarity between the different layers of a pretrained DCNN (AlexNet) and EEG signals. Similar RSA patterns were found for the generated and real EEG signals across different layers of AlexNet; representational similarity decreased as the layers increased. However, generated EEG signals showed higher similarity with the AlexNet than real EEG signals.

These findings suggest that generated EEG data based on EEG2EEG contains clearer representations of internal information in human brains than what can be obtained from real data. Thus, EEG2EEG may not only convert visual and higher-level information from individual to individual but also denoise the EEG signals.

### Training EEG2EEG model with partial data

To determine how much data is required to maximize the model's performance, we conducted several experiments where we trained our EEG2EEG converters using varying amounts of data (3000, 6000, 9000, 12000, 15000, or all 16540 samples) and compared their generated accuracies. As seen in Figure 4, the generated accuracy was greater than 60% when using 9000 samples and greater than 65% when using 12000 samples. However, the accuracy curve does not appear to have asymptoted by the full sample size, suggesting that using even more samples could lead to still better performance.

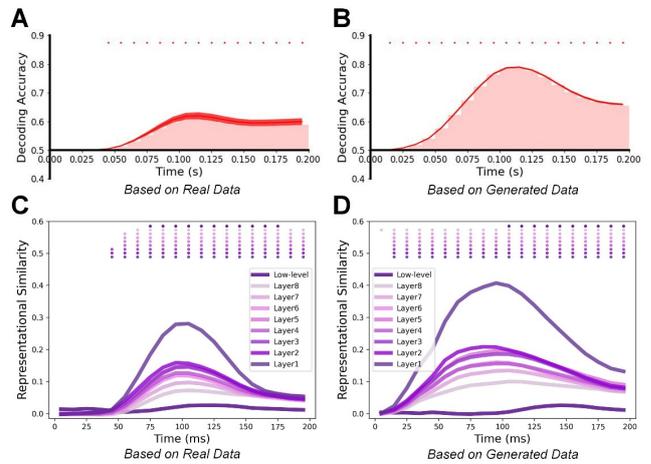

Figure 3: MVPA results. Classification-based decoding results based on (A) real data and (B) generated data. RSA results based on (C) real data and (D) generated data. Line width reflects ± SEM. Light red areas indicate clusters of time-points in which the decoding was significantly greater than chance. Color-coded small squares at the top indicate $p < 0.05$ (cluster-based permutation test).

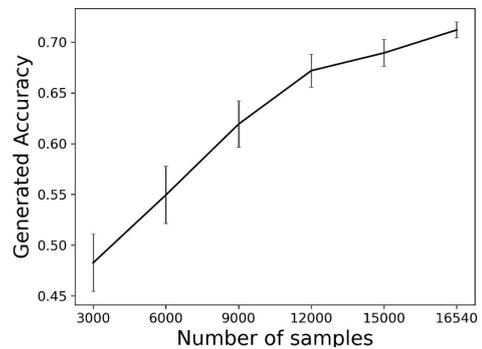

Figure 4: Generated accuracies using varying amount of data. Error bars reflect ± SEM.

## Discussion

In this study, we introduce a novel individual-to-individual EEG converter called EEG2EEG. This is the first time that a flexible EEG converter has been developed that can achieve very high conversion performance between individual subjects. Unlike previous methods, which require data from multiple subjects to find a common space of neural representations, our EEG2EEG converter has the ability to learn the mapping from one source subject to another target subject effectively. Additionally, compared to previous linear models for fMRI activity (Ho et al., 2023; Yamada et al., 2015), our EEG2EEG converter is a nonlinear model, based on a U-Net structure with full convolutional layers, and includes a loss function that uses cosine distance to align neural patterns.

Our experiments demonstrate that the EEG2EEG converter, trained on EEG signals for lots of images, can effectively generate realistic EEG signals for previously unseen visual images. The conversion performance of EEG2EEG is significantly better than that of other models tested. Additionally, the generated EEG signals based on EEG2EEG have high similarities to real EEG signals, at both ERP and single-trial level. Furthermore, MVPA results indicate that the generated data based on EEG2EEG contains sufficient and effective internal information processed by human brains as real data has. The higher decoding and RSA performance based on generated data may suggest that EEG2EEG converters also possess a certain level of signal denoising capability. The training may find the inter-subject nonlinear transformations that are more related to valid visual information than noise.

Our research suggests that our EEG2EEG converters have the ability to learn high-dimensional neural patterns across two subjects in a flexible way. And using neural converters to generate neural signals may have potential applications in EEG-based research and applications. Firstly, our EEG2EEG converters can reduce time and costs for not only ERP studies but also MVPA studies with decoding and RSA in the future. For example, we could imagine a scenario where after collecting data from multiple subjects for an initial experiment, one may only need to collect one subject's data for the new experiment and apply EEG conversion methods to generate the other subjects' data. Secondly, the EEG converters may help brain-to-brain communication in that we could use the generated signals as the reference for brain stimulation (Grau et al., 2014). Thirdly, the neural converter trained on paired data from two subjects provides us an interesting prospective to investigate individual differences. We could try to understand the mechanism of individual differences via analyzing the parameters and weights that the model learns. Furthermore, our EEG2EEG converters provide an extension of computational neuroscience and brain-computer interface models by allowing for generalization to new subjects, and they can realize real-time conversion.

One potential limitation of our current results is that they are based only on a dataset from a visual perception task, and it is unclear how well the model would generalize to other channels, types of data and tasks. For example, it will be interesting to test whether this EEG conversion framework can be applied to memory, language, and auditory tasks. If so, the potential applications of this technique could be considerable. In addition to EEG conversion, it would also be valuable to expand our nonlinear network structure to individual-to-individual fMRI conversion and explore more complex applications such as cross-modal conversion between EEG and fMRI. Neural conversion has the potential to become a powerful tool for both experimental and computational neuroscience in the future, and our study has taken a very early step in highlighting a novel EEG converter with high performance in transforming neural signals from one subject to another.

## Acknowledgements

This work was supported by research grants from the National Institutes of Health (R01-EY025648) and from the National Science Foundation (NSF 1848939).